\documentclass[12pt]{article}

\begin{document}

\begin{center}
{\bf Modified Dirac equation with Lorentz invariance violation
and its solutions for particles in an external magnetic field} \\
\vspace{5mm} S. I. Kruglov
\footnote{E-mail: krouglov@utsc.utoronto.ca}

\vspace{3mm}
\textit{Department of Chemical and Physical Sciences,\\ University of Toronto at Mississauga,\\
3359 Mississauga Rd. North, Mississauga, Ontario, Canada L5L 1C6} \\
\vspace{5mm}
\end{center}

\begin{abstract}
The second-order modified Dirac equation leading to the modified dispersion relation due to the Lorentz invariance violation corrections is suggested. The equation is formulated in the 16-component first-order form. I have obtained the projection matrix extracting solutions of the equation with definite spin projections which can be considered as the density matrix for pure spin states. Exact solutions of the equation are found for particles in the external constant and uniform magnetic field. The synchrotron radiation radius within the novel modified Dirac equation is estimated.
\end{abstract}

\section{Introduction}

The possibility of the spontaneous violation of Lorentz and CPT symmetries in the framework of the string theory was proposed in \cite{Samuel}. Thus, the Standard Model was extended and the Lorentz invariance violation (LIV) can be described by the effective field theory \cite{Colladay}. From experiment, bounds on LIV coefficients within the effective field theory were obtained \cite{Russel}. Different models of LIV in the photon sector \cite{Carroll} and fermion sector \cite{Ferreira} were investigated (there are many other publications). At present energies stringy effects are suppressed by the Planck scale $M_P=1.22\times 10^{19}$ GeV and there are not signs yet of LIV in experiments. It should be mentioned that any LIV models modify dispersion relations. It was mentioned in \cite{Amelino}, \cite{Smolin} that
quantum gravity corrections can lead to deformed dispersion relation:
\begin{equation}
p_0^2=\textbf{p}^2+m^2-\left(Lp_0\right)^\alpha\textbf{p}^2,
\label{1}
\end{equation}
where the speed of light in vacuum $c=1$, $p_0$ is an energy and $\textbf{p}$ is a momentum of a
particle and $L$ can be considered as ``minimal length" which is of the order of the Plank length $L_P=M_P^{-1}$. The subluminal propagation of particles corresponds to $L>0$ at $\alpha=1$. The last term in Eq.(1) violates the Lorentz invariance. The modified dispersion relation (1) with the parameter $\alpha=1$ was introduced in the framework of space-time foam Liouville-string models \cite{Ellis}. The wave equation for spinless particles with the dispersion equation (1) for $\alpha=1$ was considered in \cite{Ellis2}, \cite{Kruglov2}. From the analysis of the Crab Nebula synchrotron
radiation some constrains on the parameters $L$ and $\alpha$ were made \cite{Jacobson}, \cite{Ellis2}. In this paper, I postulate the modified Dirac equation for particles with spin-1/2 leading to the deformed dispersion relation (1) with $\alpha=2$. This modified Dirac equation can be considered within effective field theory with LIV in flat space-time.

The Euclidean metric and the system of units $\hbar =c=1$ is used. Greek letters run 1,2,3,4 and Latin letters run 1,2,3.

\section{Modified equation for spin-1/2 particles}

I suggest the modified Dirac equation for spin-1/2 particles of the second order:
\begin{equation}
\left(\gamma_\mu\partial_{\mu}+m-iL\gamma_4\partial_t\gamma_i\partial_i\right)\psi(x)=0,
\label{2}
\end{equation}
where $\partial_\mu=\partial/\partial x_\mu=\left(\partial/\partial x_i,\partial/(i\partial t)\right)$, $x_0=t$ is a time and notations as in \cite{Ahieser} are used. Eq.(2) is covariant under the rotational group but the additional term in Dirac equation (2) violates the invariance under the boost transformations. Thus, the Lorentz symmetry is broken and one can consider Eq.(2) as an effective wave equation with LIV introducing preferred frame effects. To obtain the dispersion relation, we investigate the plane-wave solution for positive energy $\psi(x)=\psi(p) \exp[i(\textbf{p}\textbf{x}-p_0x_0)]$ to Eq.(2). Then Eq.(2) reads
\begin{equation}
\left(i\hat{p}+m-iLp_0\gamma_4\bar{\textbf{p}}\right)\psi(p)=0,
\label{3}
\end{equation}
where $\hat{p}=\gamma_\mu p_\mu$, $\bar{\textbf{p}}=\gamma_i p_i$. One can verify, with the help of the $\gamma$-matrix algebra, that the matrix
\begin{equation}
\Lambda =i\hat{p}+m-iLp_0\gamma_4\bar{\textbf{p}}
\label{4}
\end{equation}
obeys the equation as follows:
\begin{equation}
\left(\Lambda-m\right)^2+p^2-L^2p_0^2\textbf{p}^2=0,
\label{5}
\end{equation}
where $p^2=\textbf{p}^2-p_0^2$. The matrix equation (3) possesses non-trivial solutions if det$\Lambda=0$ which is equivalent that some of the eigenvalues $\lambda$ of the matrix $\Lambda$ equal zero.
Then the requirement $\lambda=0$ leads to the modified dispersion relation:
\begin{equation}
p_0^2=\textbf{p}^2+m^2-L^2p_0^2\textbf{p}^2.
\label{6}
\end{equation}
Equation (6) corresponds to Eq.(1) for $\alpha=2$. Thus, Eq.(2) postulated realizes the modified dispersion relation (6). One can treat Eq.(2) as an effective equation for spin-1/2 particles with LIV parameter $L$ of the order to the Plank length.

\section{Wave equation in the first-order 16-component form}

Now we present the second-order equation (2) in the 16-component first-order form which is convenient for different calculations. Let us introduce, following the method of \cite{Kruglov} (see also \cite{Kruglov1}) the system of first order equations equivalences to Eq.(2)
\begin{equation}
\left(\gamma_\mu\partial_{\mu}+m\right)\psi(x)-iLm\gamma_4\partial_t\gamma_i\psi_i(x)=0,
~~~~\partial_i\psi(x)=m\psi_i(x).
\label{7}
\end{equation}
We define the 16-component wave function
\begin{equation}
\Psi (x)=\left\{ \Psi _A(x)\right\} =\left(
\begin{array}{c}
\psi(x)\\
\psi_i (x)
\end{array}
\right),
\label{8}
\end{equation}
so that  $\Psi_0(x)=\psi(x)$, $\Psi_i(x)=(1/m)\partial_i\psi(x)$.
Using the elements of the entire matrix algebra $\varepsilon
^{A,B}$, with matrix elements and products of matrices \cite{Kruglov1}
\begin{equation}
\left( \varepsilon ^{M,N}\right) _{AB}=\delta _{MA}\delta _{NB},
\hspace{0.5in}\varepsilon ^{M,A}\varepsilon ^{B,N}=\delta
_{AB}\varepsilon ^{M,N},
\label{9}
\end{equation}
where $A,B,M,N=(0,i)$, the system of equations (7), with taking into consideration Eq.(8), can be
written in the 16-component matrix form
\[
\biggl[\left(\varepsilon ^{0,0}\otimes\gamma_\mu+ mL\delta_{4\mu}\varepsilon
^{0,i}\otimes\gamma_4\gamma_i-\delta_{\mu i}\varepsilon ^{i,0}\otimes I_4\right)\partial_\mu
\]
\vspace{-8mm}
\begin{equation}
\label{10}
\end{equation}
\vspace{-8mm}
\[
+ m\left(\varepsilon ^{0,0}+\varepsilon ^{i,i}\right)\otimes I_4
\biggr]\Psi(x)=0 ,
\]
where $I_4$ is unit $4\times 4$-matrix, the $\otimes$ is the direct product of matrices, and we imply the summation over all repeated indices. One can introduce the
$16\times 16$-matrices
\begin{equation}
\Gamma_\mu=\varepsilon ^{0,0}\otimes\gamma_\mu+ mL\delta_{4\mu}\varepsilon
^{0,i}\otimes\gamma_4\gamma_i-\delta_{\mu i}\varepsilon ^{i,0}\otimes I_4,~~~
I_{16}=\left(\varepsilon ^{0,0}+\varepsilon ^{i,i}\right)\otimes I_4,
\label{11}
\end{equation}
where $I_{16}$ is unit $16\times 16$-matrix. Then Eq.(10) becomes the first-order $16\times16$-component wave equation
\begin{equation}
\left( \Gamma_\mu \partial _\mu + m\right)\Psi(x)=0 .
\label{12}
\end{equation}
From Eq.(11) we have generalized Dirac matrices as follows:
\begin{equation}
\Gamma_m=\varepsilon ^{0,0}\otimes\gamma_m-\varepsilon ^{m,0}\otimes I_4,~~~
\Gamma_4=\varepsilon ^{0,0}\otimes\gamma_4+ mL\varepsilon
^{0,i}\otimes\gamma_4\gamma_i.
\label{13}
\end{equation}
The rotational group generators in the $16$-dimension representation space are given by
\begin{equation}
J_{mn}=\left(\varepsilon^{m,n}- \varepsilon^{n,m}\right)\otimes I_4
+I_4\otimes \frac{1}{4}\left(\gamma_m\gamma_n-\gamma_n\gamma_m\right).
\label{14}
\end{equation}
The wave equation (12) is form-invariant under the rotation group transformations because matrices (13) obey the commutation relations as follows \cite{Gel'fand}:
\begin{equation}
\left[\Gamma_a , J_{m n}\right] =\delta
_{am }\Gamma_{n }-\delta _{an}\Gamma_{m},~~~~
\left[\Gamma_4,J_{m n }\right]=0. \label{15}
\end{equation}
For the positive energies $\Psi(x)=\Psi(p) \exp[i(\textbf{p}\textbf{x}-p_0x_0)]$ and Eq.(12) becomes
\begin{equation}
\left( i\check{p} + m\right) \Psi (p)=0 , \label{16}
\end{equation}
where $\check{p}=\Gamma_\mu p _\mu$. One can verify that the matrix $\check{p}$ obeys the equation
\begin{equation}
\check{p}^5 -p^2\check{p}^3+\left(mL\right)^2p^2_4\textbf{p}^2\check{p}=0 , \label{17}
\end{equation}
where $p_4=ip_0$.
Introducing the matrix of Eq.(16)
\begin{equation}
\Sigma = i\check{p} + m,
\label{18}
\end{equation}
and using Eqs.(6),(17), we obtain the matrix equation
\begin{equation}
\Sigma\left(\Sigma-m\right)\left(\Sigma-2m\right)\left[\left(\Sigma-m\right)^2+p^2+m^2\right]
=0. \label{19}
\end{equation}
With the help of Eq.(19), we find solutions to Eq.(16) in the form of the projection matrix
\begin{equation}
\Pi=N\left(\Sigma-m\right)\left(\Sigma-2m\right)\left[\left(\Sigma-m\right)^2+p^2+m^2\right].
 \label{20}
\end{equation}
Indeed, it follows from Eq.(19) that $\left(i\check{p} + m\right)\Pi=0$. Therefore, every column of the matrix $\Pi$ is the solution to Eq.(16). The normalization
constant $N$ can be find from the requirement \cite{Fedorov}:
\begin{equation}
\Pi^2=\Pi. \label{21}
\end{equation}
From Eqs.(19)(20), after some calculations, one obtains the normalization constant
\begin{equation}
N=\frac{1}{2m^2\left(p^2+2m^2\right)}.
\label{22}
\end{equation}
We treat the matrix $\Pi$ as a density matrix which describes impure spin states.
To extract pure spin states, one may introduce the spin projection operator
\begin{equation}
\sigma_p=-\frac{i}{2|\textbf{p}|}\epsilon_{abc}\textbf{p}_a
J_{bc},
\label{23}
\end{equation}
where $|\textbf{p}| =\sqrt{p_i^2}$, which obeys the minimal matrix equation as follows \cite{Kruglov3}:
\begin{equation}
\left(\sigma_p^2-\frac{1}{4}\right)\left(\sigma_p^2-\frac{9}{4}\right)=0.
 \label{24}
\end{equation}
With the help of the method \cite{Fedorov}, we find
from Eq.(24) the projection operators extracting spin projections $\pm 1/2$:
\begin{equation}
P_{\pm 1/2}=\mp\frac{1}{2}\left(\sigma_p\pm\frac{1}{2}
\right)\left(\sigma_p^2-\frac{9}{4}\right).
 \label{25}
\end{equation}
Then the projection operator extracting pure spin states with positive energy is
\begin{equation}
\Delta_{\pm 1/2}= \Pi P_{\pm 1/2}, \label{26}
\end{equation}
obeying $\Delta_{\pm 1/2}^2=\Delta_{\pm 1/2}$.
Thus, $\Delta_{\pm 1/2}$ may be explored in calculations of some processes including electrons with taking into consideration LIV parameter $L$.

\section{Spin-1/2 particle in an external magnetic field}

Let us consider a particle in an external uniform and static magnetic field along the $x_3$ axis, $\textbf{H}=(0,0,H)$. Then the 4-potential can be chosen as
\begin{equation}
A_\mu=\left(0,Hx_1,0,0\right).
\label{27}
\end{equation}
The electromagnetic interaction of particles in Eq.(2) can be introduced by the standard substitution $\partial_\mu\rightarrow D_\mu=\partial_\mu-ieA_\mu$. For the electrons $e=-e_0$, $e_0>0$ and Eq.(2)  becomes
\[
\biggl[\gamma_m\partial_m+ie_0Hx_1\gamma_2-i\gamma_4\biggl(1+L\gamma_m\partial_m
\]
\vspace{-7mm}
\begin{equation} \label{28}
\end{equation}
\vspace{-7mm}
\[
+ie_0HLx_1\gamma_2\biggr)\partial_t+m \biggr]\Psi\left(x\right)=0.
\]
One can look for a solution to Eq.(28), as for Dirac equation \cite{Ahieser}, \cite{Sokolov}, in the form
\begin{equation}
\Psi\left(x\right)=\frac{1}{\sqrt{L_2L_3}}\exp\left[i\left(p_2x_2+p_3x_3-p_0t\right)\right]\Psi(x_1),
\label{29}
\end{equation}
where $p_2=2\pi n_2/L_2$, $p_3=2\pi n_3/L_3$; $n_2$ and $n_3$ are integer quantum numbers.
Taking into account Eq.(29), equation (28) becomes
\[
\biggl\{\left(\gamma_1+iLp_0\alpha_1\right)\partial_1+i\left[\gamma_2p_2+\gamma_3p_3+iLp_0\left(
\alpha_2p_2+\alpha_3p_3\right)\right]
\]
\vspace{-7mm}
\begin{equation} \label{30}
\end{equation}
\vspace{-7mm}
\[
+ie_0Hx_1\left(\gamma_2+iLp_0\alpha_2\right)+m\biggr\}\Psi\left(x_1\right)=p_0\gamma_4\Psi\left(x_1\right)
,
\]
where we have introduced matrices as follows \cite{Ahieser}:
\begin{equation}
\alpha_k=i\gamma_4\gamma_k=\left(
\begin{array}{cc}
0&\sigma_k\\
\sigma_k&
\end{array}
\right)
,~~~~
\gamma_4=\left(
\begin{array}{cc}
I_2&0\\
0&-I_2
\end{array}
\right),
\label{31}
\end{equation}
and $\sigma_k$ are the Pauli matrices. Replacing the bispinor wave function
\begin{equation}
\Psi\left(x_1\right)=\left(
\begin{array}{c}
\varphi(x_1)\\
\chi(x_1)
\end{array}
\right),
\label{32}
\end{equation}
into Eq.(30) with the help of Eq.(31), we obtain the system of equations
\[
\left(1-Lp_0\right)\left(-i\sigma_1\partial_1+p_2\sigma_2+p_3\sigma_3
+e_0Hx_1\sigma_2\right)\chi=\left(p_0-m\right)\varphi,
\]
\vspace{-8mm}
\begin{equation}
\label{33}
\end{equation}
\vspace{-8mm}
\[
\left(1+Lp_0\right)\left(i\sigma_1\partial_1-p_2\sigma_2-p_3\sigma_3
-e_0Hx_1\sigma_2\right)\varphi=-\left(p_0+m\right)\chi.
\]
From Eqs.(33) one finds the equation for $\varphi$:
\begin{equation}
\left[\left(1-L^2p^2_0\right)\left(i\sigma_1\partial_1-p_2\sigma_2-p_3\sigma_3
-e_0Hx_1\sigma_2\right)^2+m^2-p_0^2\right]\varphi(x_1)=0.
\label{34}
\end{equation}
To take into consideration the spin projection of an electron, we require that the $\varphi$ is the eigenfunction of the operator of the spin projection on the magnetic field direction $\sigma_3$ \cite{Ahieser}:
\begin{equation}
\sigma_3\varphi(x_1,\mu)=\mu\varphi(x_1,\mu),
\label{35}
\end{equation}
where $\mu=\pm 1$. Then Eq.(34), with the help of Eq.(35) and the properties of Pauli's matrices, becomes   \begin{equation}
\left[-\partial_1^2+p_3^2+\left(p_2+e_0Hx_1\right)^2+e_0H\mu\right]\varphi(x_1,\mu)=
\frac{p_0^2-m^2}{1-L^2p_0^2}\varphi(x_1,\mu).
\label{36}
\end{equation}
After introducing new variable
\[
\xi=\sqrt{e_0H}\left(x_1+\frac{p_2}{e_0H}\right)
\]
Eq.(36) takes the form of the equation for the harmonic oscillator:
\begin{equation}
\left(-\frac{d^2}{d\xi^2}+\xi^2\right)\varphi(\xi,\mu)=
\frac{1}{e_0H}\left(\frac{p_0^2-m^2}{1-L^2p_0^2}-p_3^2-e_0H\mu\right)\varphi(\xi,\mu).
\label{37}
\end{equation}
It should be mentioned that $p_2$ defines the coordinate of the orbit center \cite{Sokolov} $x_1=-p_2/(e_0H)$ ($\xi=0$).
The requirement that $\varphi(\xi,\mu)\rightarrow 0$ at $\xi\rightarrow \pm\infty$ gives
\begin{equation}
\frac{1}{e_0H}\left(\frac{p_0^2-m^2}{1-L^2p_0^2}-p_3^2-e_0H\mu\right)=2n+1,
\label{38}
\end{equation}
where $n=0,1,2,...$ is the principal quantum number.
From Eq.(38) one obtains the energy quantization:
\begin{equation}
p_0^2= m^2+\left(1-L^2p_0^2\right)\left[p_3^2+e_0H\left(2n+1+\mu\right)\right].
\label{39}
\end{equation}
Due to the condition (39) the solution to Eq.(37) is given by
\begin{equation}
\varphi(\xi)=C\exp\left(-\frac{\xi^2}{2}\right)H_n\left(\xi\right),
\label{40}
\end{equation}
where $H_n\left(\xi\right)$ are the Hermite polynomials and $C$ is the normalization constant.
Eq.(39) shows that the energy of the electron depends on the LIV parameter $L$. At $L=0$ one arrives to the well-known result \cite{Ahieser}. From (29), (32), (33), one may find the bispinor wave function $\Psi(x)$. The normalization constant $C$ is defined from the relation \cite{Ahieser}:
\begin{equation}
\int_0^{L_3}dx_3 \int_0^{L_2}dx_2\int_{-\infty}^\infty \Psi^+(x)\Psi(x)dx_1 =1,
\label{41}
\end{equation}
where $\Psi^+(x)$ is the Hermitian conjugate function.
One can apply Eqs.(29), (32), (40) to study the synchrotron radiation of electrons with modified dispersion relation (39). Electrons in an external magnetic field moves in helical orbits and emit the synchrotron radiation and the frequency depends on the orbit radius \cite{Sokolov}. Let us consider the high (ultra-relativistic) energies of electrons ($p_0\gg m$). In the case $p_3=0$ particles move in a plane and at $2L^2e_0Hn\ll 1$ Eq.(39) is approximated as
\begin{equation}
p_0\approx \sqrt{2e_0Hn}-\sqrt{2}L^2\left(e_0Hn\right)^{3/2}.
\label{42}
\end{equation}
The classical radius of the orbit is given by \cite{Sokolov}:
\begin{equation}
R =\frac{\beta p_0}{e_0H},
\label{43}
\end{equation}
with $\beta=v$ being the electron velocity. Substituting Eq.(42) into (43) at $v\approx 1$, we obtain the orbit radius
\begin{equation}
R \approx R_0-L^2\sqrt{2e_0H}n^{3/2},
\label{44}
\end{equation}
where
\begin{equation}
R_0 \approx\sqrt{\frac{2n}{e_0H}}
\label{45}
\end{equation}
is the orbit radius in the framework of the Klein$-$Gordon equation \cite{Sokolov}.
It follows from Eq.(44) that LIV parameter $L$ reduces the orbit radius. The angular orbital frequency $\omega=v/R$ becomes greater due to the Lorentz-violating term.

To estimate the contribution of LIV parameter $L$ to observable, we use the data from experiments with magnetic trapping of electrons \cite{Dehmelt}. Electrons in the Penning trap 90$\%$ of the time are in the cyclotron ground state $n=1$ and in external magnetic field $H=6$ T. Using the Heaviside$-$Lorentz system with the fine structure constant $\alpha=e^2/(4\pi)$ and the SI units which are related to the energy units $1$ T$=195.5$ eV$^2$, $1$ m$=5.1\times10^6$ eV$^{-1}$, we obtain the radius change at $L=L_P$ from Eq.(44): $\Delta R=L_P^2\sqrt{2e_0H}n^{3/2}\approx 10^{-60}$ m. Thus, when the LIV parameter $L$ is equal to the Planck length $L_P$ the change of the radius of the electron orbit is extremely small and Penning trap experiments are not relevant.

From the Crab Nebula data, one can use the possible energy of electrons $p_0=1$ TeV, and the magnetic field $H=260~\mu$G . We obtain the principal quantum number $n\approx p_0^2/(2e_0H) \approx 3\times 10^{29}$, and the change of the synchrotron radius at $L=L_P$, $\Delta R=L_P^2\sqrt{2e_0H}n^{3/2}\approx
10^{-22}$ m. This value is very small compared to the classical radius (45). If the LIV parameter $L$ is much greater than the Planck length $L_P$, then the LIV effect has to be taken into account.

\section{Conclusion}

The wave equation for spin-1/2 particles (electrons) suggested leads to the modified dispersion relation (1) with $\alpha =2$ which was discussed in the context of quantum gravity corrections in \cite{Ellis2}. I have formulated the novel equation in the first-order form and found the minimal polynomial of matrices of the equation and the spin projection operators. This allowed us to obtain the projection matrix extracting solutions for free particles in the momentum space. One may consider the projection matrix found as the density matrix with taking into account LIV effects. Such density matrix can be explored in different calculations of quantum possesses. We have found exact solutions to the wave equation for spin-1/2 particles in a constant and uniform external magnetic field. Exact solutions obtained can be used for the analysis of the synchrotron radiation of electrons with taking into consideration LIV corrections. The synchrotron radius corrections leading to increasing the synchrotron frequency were estimated. We leave the analysis of bounds on the LIV parameter $L$ from astrophysical data for further investigations.


\begin{thebibliography}{99}

\bibitem{Samuel} V. A. Kostelecky and S. Samuel, Phys. Rev. \textbf{D39}, 683 (1989); Phys. Rev. Lett. \textbf{63}, 224 (1989); ibid \textbf{66}, 1811 (1991); V. A. Kostelecky and R. Potting, Nucl. Phys. \textbf{B359}, 545 (1991); Phys. Lett. \textbf{B381}, 89 (1996) [hep-th/9605088 [hep-th]]; Phys. Rev. \textbf{D51}, 3923 (1995) [hep-ph/9501341 [hep-ph]].

\bibitem{Colladay} D. Colladay and V. A. Kostelecky, Phys. Rev. \textbf{D55}, 6760 (1997)
[hep-ph/9703464]; ibid \textbf{D58}, 116002 (1998) [hep-ph/9809521];
S. R. Coleman and S. L. Glashow, Phys. Rev. \textbf{D59}, 116008 (1999) [hep-ph/9812418].

\bibitem{Russel} V. A. Kostelecky and N. Russell, Rev. Mod. Phys. \textbf{83}, 11 (2011) [arXiv:0801.0287 [hep-ph]].

\bibitem{Carroll} S. M. Carroll, G. B. Field and R. Jackiw, Phys. Rev. \textbf{D41}, 1231 (1990);
R. Jackiw and V. A. Kostelecky, Phys. Rev. Lett. 82, 3572 (1999) [hep-ph/9901358 [hep-ph]];
S. I. Kruglov, Phys. Lett. \textbf{B652}, 146 (2007) [arXiv:0705.0133 [hep-ph]]; V. A. Kostelecky and M. Mewes, Phys. Rev. \textbf{D80}, 015020 (2009) [arXiv:0905.0031 [hep-ph]].

\bibitem{Ferreira} M. Chaichian, W. F. Chen and R. Gonzalez Felipe, Phys. Lett. \textbf{B503}, 215 (2001)
[hep-th/0010129 [hep-th]]; D. V. Ahluwalia, Class. Quant. Grav. \textbf{22}, 1433 (2005)
[hep-th/0503141 [hep-th]]; M. M. Ferreira, Jr. and F. M. O. Moucherek, Int. J. Mod. Phys. \textbf{A21},
6211 (2006) [hep-th/0601018 [hep-th]]; A. E. Bernardini, Phys. Rev. \textbf{D75}, 097901 (2007)
[arXiv:0706.3932 [hep-ph]]; P. A. Bolokhov and M. Pospelov, Phys. Rev. \textbf{D77}, 025022 (2008)
[hep-ph/0703291 [hep-ph]]; H. Belich et al, Eur. Phys. J. \textbf{C62}, 425 (2009) [arXiv:0806.1253 [hep-th]]; B. Goncalves,
Y. N. Obukhov and I. L. Shapiro. Phys. Rev. \textbf{D80}, 125034 (2009) [arXiv:0908.0437 [hep-th]];
A. Kostelecky and M. Mewes, Phys. Rev. \textbf{D85}, 096005 (2012) [arXiv:1112.6395 [hep-ph]].

\bibitem{Amelino} G. Amelino-Camelia, Int. J. Mod. Phys. \textbf{D11}, 35 (2002) [arXiv:gr-qc/0012051];
 New J. Phys. \textbf{6}, 188 (2004) [arXiv:gr-qc/0212002].

\bibitem{Smolin} J. Magueijo and L. Smolin, Phys. Rev. Lett., \textbf{88}, 190403 (2002)
[arXiv:hep-th/0112090]; Phys. Rev. \textbf{D67}, 044017 (2003) [arXiv:gr-qc/0207085].

\bibitem{Ellis} J. R. Ellis, N. E. Mavromatos and D. V. Nanopoulos, Phys. Lett. \textbf{B293},  37 (1992) [arXiv:hep-th/9207103];
 Chaos, Solitons and Fractals \textbf{10}, 345 (1999) [arXiv:hep-th/9805120].

\bibitem{Ellis2} J. Ellis, N. E. Mavromatos and A. S. Sakharov, Astropart. Phys. \textbf{20}, 669 (2004)
[arXiv:astro-ph/0308403].

\bibitem{Kruglov2} S. I. Kruglov, Modified wave equation for spinless particles and its solutions in an external magnetic field, arXiv:1207.6573 [hep-th].

\bibitem{Jacobson} T. Jacobson, S. Liberati and D. Mattingly, Nature \textbf{424}, 1019 (2003)
[arXiv:astro-ph/0212190].

\bibitem{Ahieser}  A. I. Ahieser and V. B. Berestetskii, Quantum
Electrodynamics (New York: Wiley Interscience, 1969).

\bibitem{Kruglov} S. I. Kruglov, Int. J. Mod. Phys. \textbf{A27}, 1250081 (2012)
[arXiv:1201.2391  [hep-ph]].

\bibitem{Kruglov1} S. I. Kruglov, Symmetry and Electromagnetic Interaction of Fields with
Multi-Spin (Nova Science Publishers, Huntington, New York, (2001)).

\bibitem{Gel'fand} I. M. Gel'fand, R. A. Minlos and Z. Ya. Shapiro, Representations
of the Rotation and Lorentz Groups and their Applications
(Pergamon, New York, 1963).

\bibitem{Fedorov} F. I. Fedorov,  Sov. Phys. - JETP \textbf{35}(8), 339 (1959) (Zh.
Eksp. Teor. Fiz. \textbf{35}, 493 (1958)).

\bibitem{Kruglov3} S. I. Kruglov, Int. J. Mod. Phys. \textbf{A21}, 1143 (2006)
[hep-th/0405088 [hep-th]].

\bibitem{Sokolov} A. A. Sokolov and I. M. Ternov, Radiation from Relativistic electrons, edited by C. W. Kilmister, AIP translation Series (1986).

\bibitem{Dehmelt} L. S. Brown and G. Gabrielse, Rev. Mod. Phys. \textbf{58}, 233 (1986);
L. S. Brown et al, Phys. Rev. \textbf{A37}, 4163 (1988);
H. Dehmelt et al, Phys. Rev. Lett. \textbf{83}, 4694 (1999) [hep-ph/9906262].

\end{thebibliography}
\end{document}